\pdfoutput=1

\documentclass[%
 reprint,
 superscriptaddress,
 nobibnotes,
 amsmath,amssymb,
 aps,
 showkeys,
 longbibliography,
]{revtex4-2}

\usepackage{textcomp, gensymb}
\usepackage{graphicx}
\usepackage{dcolumn}
\usepackage{bm}
\usepackage{chemformula}
\usepackage[separate-uncertainty=true,separate-uncertainty,multi-part-units=single]{siunitx}
\usepackage{hyperref}
\hypersetup{colorlinks=true,allcolors=magenta}
\usepackage{hyperref}

\usepackage{footnote}

\begin{document}

\title{Computing solubility and thermodynamics properties of H$_2$O$_2$ in water}
\email[]{Supplementary Information available}
\author{Tijin H.G. Saji}
\affiliation{Department of Applied Physics and Science Education, Technical University of Eindhoven, PO Box 513,
5600 MB, Eindhoven, The Netherlands.}

\affiliation{Institute for Complex Molecular Systems, PO Box 513,
5600 MB, Eindhoven, The Netherlands.}
\author{José Manuel Vicent-Luna}
\affiliation{Department of Applied Physics and Science Education, Technical University of Eindhoven, PO Box 513,
5600 MB, Eindhoven, The Netherlands.}
\author{Thijs J.H. Vlugt}
\affiliation{Process $\&$ Energy Department, Faculty of Mechanical, Maritime and Materials Engineering, Delft University of Technology, Leeghwaterstraat 39, 2628 CB, Delft, The Netherlands.}
\author{Sofía Calero}
\email[Corresponding authors: ]{s.calero@tue.nl, b.bagheri@tue.nl}
\affiliation{Department of Applied Physics and Science Education, Technical University of Eindhoven, PO Box 513,
5600 MB, Eindhoven, The Netherlands.}
\author{Behnaz Bagheri}
\email[Corresponding authors: ]{s.calero@tue.nl, b.bagheri@tue.nl}
\affiliation{Department of Applied Physics and Science Education, Technical University of Eindhoven, PO Box 513,
5600 MB, Eindhoven, The Netherlands.}

\affiliation{Institute for Complex Molecular Systems, PO Box 513,
5600 MB, Eindhoven, The Netherlands.}



\date{\today}

\begin{abstract}
Hydrogen peroxide plays a key role in many environmental and industrial chemical processes. We performed classical Molecular Dynamics and Continuous Fractional Component Monte Carlo simulations to calculate thermodynamic properties of H\textsubscript{2}O\textsubscript{2} in aqueous solutions. The quality of the available force fields for H$_2$O$_2$ developed by Orabi \& English, and by Cordeiro was systematically evaluated. To assess which water force field is suitable for predicting properties of H$_2$O$_2$ in aqueous solutions, four water force fields were used, namely the TIP3P, TIP4P/2005, TIP5P-E, and a modified TIP3P force field. While the computed densities of pure H$_2$O$_2$ in the temperature range of 253~-~353~K using the force field by Orabi \& English are in excellent agreement with experimental results, the densities using the force field by Cordeiro are underestimated by 3$\%$. The TIP4P/2005 force field in combination with the H$_2$O$_2$ force field developed by Orabi \& English can predict the densities of H$_2$O$_2$ aqueous solution for the whole range of H$_2$O$_2$ mole fractions in very good agreement with experimental results. The TIP4P/2005 force field in combination with either of the H$_2$O$_2$ force fields can predict the viscosities of H$_2$O$_2$ aqueous solutions for the whole range of H$_2$O$_2$ mole fractions in reasonably good agreement with experimental results. The computed diffusion coefficients for H$_2$O$_2$ and water molecules using the TIP4P/2005 force field with either of the H$_2$O$_2$ force fields are almost constant for the whole range of H$_2$O$_2$ mole fractions. Hydrogen bond analysis showed a steady increase in the number of hydrogen bonds with the solute concentrations in H$_2$O$_2$ aqueous solutions for all combinations except for the Cordeiro-TIP5P-E and Orabi-TIP5P-E systems, which showed a minimum at intermediate concentrations. The Cordeiro force field for H$_2$O$_2$ in combination with either of the  water force fields can predict the Henry coefficients of H$_2$O$_2$ in water in better agreement with experimental values than the force field by Orabi \& English. 

\end{abstract}


\maketitle

\section{Introduction}
\label{sec:intro}
Hydrogen peroxide, H$_2$O$_2$, has attracted considerable interest as it plays a key role in the oxidative chemistry of the troposphere. It can be found both in the gas and in the aqueous phase~\cite{lelieveld1990influences,crutzen1999background}, and has several industrial~\cite{lewis2019recent}, environmental~\cite{chen2023photocatalytic}, and biological~\cite{wang2017leveraging} applications. The recombination of hydroperoxyl~(HO$_2$) radicals is the most important chemical pathway leading to the production of H$_2$O$_2$ in the troposphere~\cite{moller2009atmospheric,vione2003atmospheric,lee2000hydrogen}. Subsequently, H$_2$O$_2$ can lead to the acidification of clouds, rain, and fog by oxidizing SO$_2$ and converting it into H$_2$SO$_4$ (and to a less extent oxidizing NO$_2$ and converting it into HNO$_3$)~\cite{hoffmann1975kinetics,penkett1979importance,martin1981aqueous,damschen1983aqueous,calvert1985chemical}. H$_2$O$_2$ also serves as a reservoir of HO$_{\text x}$ radicals that are key oxidants in controlling the self-cleaning of the atmosphere~\cite{kleinman1991seasonal,logan1981tropospheric,levy1971normal}. 

H$_2$O$_2$ was first synthesized by Thenard~\cite{fierro2006hydrogen} by the reaction of barium peroxide with nitric acids in 1818 and is now considered an important reagent of green chemistry since it decomposes to water and oxygen as the only reaction products. 
This feature makes H$_2$O$_2$ an environmentally friendly oxidizing agent for a wide range of applications such as pulp and paper bleaching, textile applications, detergent applications, disinfectant applications, wastewater treatment, and chemical oxidation processes~\cite{campos2006hydrogen,hou2020production}. It could also serve as a liquid fuel, an alternative to H$_2$ and O$_2$, in a fuel cell ~\cite{mase2016seawater,kofuji2016carbon,fukuzumi2017production}.

H$_2$O$_2$ is currently produced on an industrial scale with the anthraquinone oxidation (AO) process in which hydrogen, atmospheric oxygen, and an anthraquinone derivative~(typically 2-alkyl-anthraquinone) are used with the latter acting as a reaction carrier~\cite{campos2006hydrogen,hou2020production}.  The ubiquitous AO process involves multiple steps which require significant energy input and generates waste. In addition, the transport, storage, and handling of bulk H$_2$O$_2$ involve hazards as it is irritating to nose and eyes, and high concentration of H$_2$O$_2$ is explosive\cite{shanley1947highly}. Other methods for large-scale production of H$_2$O$_2$ include partial oxidation of primary or secondary alcohols, and electrochemical methods~\cite{perry2021future}. Novel alternatives are under investigation such as direct synthesis of H$_2$O$_2$ from O$_2$ and H$_2$ using a variety of catalysts like alumina, silica, carbon, solvents~(e.g., water)~\cite{samanta2008direct,burch2003investigation}, photocatalytic reactions over semiconductors where reactive oxygen-based species~(e.g., OH$^{\bullet}$, O$^{2-}$, and H$_2$O$_2$) are formed at the surface of semiconductor oxides under UV irradiation~\cite{genti2003tubular}. An alternative technology to produce H$_2$O$_2$ is to use low temperature~(or non-thermal) plasmas~\cite{bruggeman_non-thermal_2009,bruggeman2017foundations} which allows H$_2$O$_2$ production 
at ambient temperatures and pressures~\cite{joshi1995formation,locke2011review,yue2016measurements,suresh2022production,cameli2022modular}. This enables direct delivery of H$_2$O$_2$ to different substrates; even to heat sensitive substrates such as living tissues. The latter has led to biomedical applications of low temperature plasmas\cite{bruggeman2016plasma}. For such applications, it is important to know which mechanisms determine the uptake of plasma products (e.g., H$_2$O$_2$) in the liquid around the cells. For this, information on solubility and thermodynamics properties of plasma products are necessary so that this can be leveraged into macroscopic plasma fluid models\cite{tian2014atmospheric} to predict the final concentration of plasma products in the liquid phase. The motivation of this work is to provide such data for H$_2$O$_2$ as limited data are available.

Due to the pivotal role of H$_2$O$_2$ in many chemical processes, many experimental and computational studies have been conducted to investigate its properties. The crystal structure of H\textsubscript{2}O\textsubscript{2} was investigated using diffraction methods or Raman spectroscopy in Refs.~\cite{giguere1943electron, giguere1974raman, abrahams1951crystal, savariault1980experimental, fritchie1981neutron}. Other experimental studies have investigated its densities~\cite{maass1920properties}, viscosities ~\cite{phibbs1951hydrogen}, vibrational spectra~\cite{redington1962studies, giguere1974raman, koput1986r0, khachkuruzov1978molecular,hunt1965internal}, vapor pressures~\cite{maass1924properties} and other thermodynamics properties~\cite{foley1951hydrogen}. In addition, densities, freezing points, and vapor pressures of aqueous H\textsubscript{2}O\textsubscript{2} solutions were investigated experimentally in Refs.~\cite{maass1920properties, easton_behaviour_1952,foley1951hydrogen2,scatchard1952vapor}.

Various computational studies have been carried out which shed light on structural properties of H\textsubscript{2}O\textsubscript{2} monomers as well as its clusters, torsional barrier energies, and vibrational-rotational energy levels~\cite{cremer1978theoretical,koput1998potential,kulkarni2003ab,elango2006hydrogen,kulkarni2005structures} using quantum mechanical approaches. Structure and dynamics of H\textsubscript{2}O\textsubscript{2} in water were also investigated using quantum mechanical methods in Refs.~\cite{moin2012ab, ferreira2011electronic,priyadarsini_structure_2022}. 

In this work, we use force field based Molecular Dynamics~(MD) and Continuous Fractional Component Monte Carlo~(CFCMC) simulations with the purpose of obtaining solubilities and thermodynamics properties of H$_2$O$_2$ in water, for the first time, in a systematic manner such that the quality of the available force fields for H$_2$O$_2$ are assessed.  

 Although several force fields are available for H\textsubscript{2}O\textsubscript{2}~\cite{dominguez_catalase_2010, chung2007diffusion, dominguez_how_2014, vacha2004adsorption}, only a few of them have been parameterized with respect to the interactions between both H\textsubscript{2}O\textsubscript{2}~-~H\textsubscript{2}O\textsubscript{2} and H\textsubscript{2}O\textsubscript{2}~-~H\textsubscript{2}O. One is the ABEEM/MM, the atom-bond
 electronegativity equalization fluctuating charge molecular
 force field~\cite{yang1997atom,wang1999atom}, 
 which is computationally very expensive due to its complex potential energy functional form~\cite{yang1997atom, wang1999atom}. A simple additive potential model for  H\textsubscript{2}O\textsubscript{2} was proposed by Orabi $\&$ English~\cite{orabi_simple_2018} which was parameterized to account for interactions of  H\textsubscript{2}O\textsubscript{2} with itself and with water. The model was calibrated with regard to the experimental density and heat of vaporization of pure liquid H\textsubscript{2}O\textsubscript{2} at 0\degree~C, and was able to reproduce the experimental diffusion coefficient at 0\degree~C and the heat capacity at 25\degree~C of liquid H\textsubscript{2}O\textsubscript{2}. With a combination of the modified TIP3P water force field~\cite{neria1996simulation}, the H\textsubscript{2}O\textsubscript{2} force field could predict the experimental hydration free energies and densities of aqueous H\textsubscript{2}O\textsubscript{2} solutions~\cite{orabi_simple_2018}. 
 Another force field parametrization 
 is from the work of Cordeiro~\cite{cordeiro_reactive_2014}, wherein the bonded interactions were obtained from \textit{ab intio} quantum calculations~\cite{de2001density, koput1998potential}, and the Lennard-Jones parameters and partial charges were modified to reproduce the properties of pure liquid H\textsubscript{2}O\textsubscript{2} and its hydration free energy. This force field was used to study the distribution, mobility and residence times of H\textsubscript{2}O\textsubscript{2} at the interface of water and phospholipid biomembranes. In addition, there is another parametrization in the paper by Vácha \textit{et al.}~\cite{vacha2004adsorption} in which the behaviour of H\textsubscript{2}O\textsubscript{2} at the air-water interface was investigated. The force field by Vácha \textit{et al.}~\cite{vacha2004adsorption} is a rigid force field; it only includes electrostatic and van der Waals interactions which were calibrated against the experimental hydration energies of H$_2$O$_2$. 
 

In this manuscript, we evaluate the quality of the force fields which were developed by Cordeiro~\cite{cordeiro_reactive_2014}, and Orabi $\&$ English~\cite{orabi_simple_2018} for predicting the thermodynamic properties of H$_2$O$_2$ in aqueous solutions. Both force fields by Cordeiro~\cite{cordeiro_reactive_2014}, Orabi \& English~\cite{orabi_simple_2018} are non-rigid, thereby we exclude the force field by Vácha \textit{et al.}~\cite{vacha2004adsorption} from our study as it is a rigid force field. We compute the densities of pure H\textsubscript{2}O\textsubscript{2} for a range of temperatures~(253~K to 353~K), and compare the results with experimental values. In addition, we compute densities, viscosities, and diffusion coefficients of H$_2$O$_2$ and water in aqueous solutions of H$_2$O$_2$ for the whole range of H\textsubscript{2}O\textsubscript{2} mole fractions at ambient temperatures and pressures. To evaluate which water force field is suitable for predicting properties of H\textsubscript{2}O\textsubscript{2} aqueous solutions, we use four different water force fields: TIP3P~\cite{jorgensen1983comparison,neria1996simulation} as it performs better in calculating the specific heats of water~\cite{mao2012thermal}, TIP5P-E as it can capture the thermal conductivities of water ~\cite{mao2012thermal} and TIP4P/2005~\cite{abascal2005general} as it can predict the densities and self-diffusion coefficients of water with commendable accuracy~\cite{abascal2005general}. The fourth water force field is a modified version of TIP3P~(mTIP3P)~\cite{neria1996simulation}, which was used in the work by Orabi $\&$ English~\cite{orabi_simple_2018}. The results are compared with experimental values. Finally, we compute the Henry coefficients of H$_2$O$_2$ in water at 300~K.

The rest of this manuscript is organized as follows. In section~\ref{Methods}, details of the force fields which were developed by Cordeiro~\cite{cordeiro_reactive_2014}, and Orabi $\&$ English~\cite{orabi_simple_2018} are provided, and the MD and CFCMC simulations are described. The results are presented and discussed in section~\ref{FF validation}. Finally, concluding remarks are presented in section~\ref{Conclusion}. 
\section{Methodology} \label{Methods}
\subsection{Force Fields} \label{FFs}
\begin{table*}
\small
  \caption{\ Potential energy functions for force fields developed by Cordeiro~\cite{cordeiro_reactive_2014}, and Orabi $\&$ English~\cite{orabi_simple_2018}. $E_{\text{{bonds}}}$, $E_\text{{angles}}$, $E_\text{{dihedrals}}$, $E_\text{{vdW}}$, and $E_\text{{electrostatic}}$ represent the stretching, bending, torsional, van der Waals and electrostatic energies, respectively. The definition of parameters is explained in the text~(see section~\ref{FFs}). The parameters are provided in Tables S2 and S3 of the Supplementary Information.} 
  \label{tbl:forcefields}
  \begin{tabular*}{\textwidth}{@{\extracolsep{\fill}}llllllll}
    \hline
    Force field & $E_{\text{{bonds}}}$ & $E_\text{{angle}}$& $E_\text{{dihedrals}}$ & $E_\text{{vdW}}$ & $ E_\text{{electrostatic}}$\\
    \hline
    Cordeiro~\cite{cordeiro_reactive_2014}  & $\frac{1}{4}k_{b}(b\textsuperscript{2}-b\textsubscript{0}\textsuperscript{2})\textsuperscript{2}$ & $\frac{1}{2}k_{\theta}(\cos{\theta}-\cos{\theta}_{0})^{2}$  & ${\sum_{n=0}^{5}}~C_{n}(\cos{\psi})^{n}$ & $4\epsilon_{ij}[(\frac{\sigma_{ij}}{r_{ij}})^{12}-(\frac{\sigma_{ij}}{r_{ij}})^{6}]$ & $\frac{1}{4\pi\epsilon_{o}}\frac{q_{i}q_{j}}{r_{ij}}$ \\
    Orabi $\&$ English~\cite{orabi_simple_2018} & $\frac{1}{2}k_{b}(b-b_{0})^2 $ & $\frac{1}{2}k_{\theta}(\theta-\theta_{0})^2 $& $k_{\phi}(1+\cos{2\phi-\delta})$ & $4\epsilon_{ij}[(\frac{\sigma_{ij}}{r_{ij}})^{12}-(\frac{\sigma_{ij}}{r_{ij}})^{6}]$ &  $\frac{1}{4\pi\epsilon_{o}}\frac{q_{i}q_{j}}{r_{ij}}$ \\
     \\
    \hline
  \end{tabular*}
\end{table*}

Both force fields developed by Cordeiro~\cite{cordeiro_reactive_2014}, and Orabi $\&$ English~\cite{orabi_simple_2018} for H\textsubscript{2}O\textsubscript{2} consider a non-rigid H\textsubscript{2}O\textsubscript{2} molecule, that is, they incorporate bonds, angles and dihedrals information with the van der Waals~(vdW) and electrostatic interactions. The total potential energy (\textit{E}\textsubscript{total}) is given by  
\begin{equation}
    E_{\text{total}} = E_{\text{{bonds}}} + E_\text{{angles}} + E_\text{{dihedrals}} + E_\text{{vdW}} + E_\text{{electrostatic}},
\end{equation}
where $E_{\text{{bonds}}}$, $E_\text{{angles}}$, $E_\text{{dihedrals}}$, $E_\text{{vdW}}$, and $E_\text{{electrostatic}}$ are presented in Table~\ref{tbl:forcefields} for both force fields. The bonded interaction~parameters ($E_{\text{{bonds}}}$, $E_\text{{angles}}$, and $E_\text{{dihedrals}}$) listed in Table~\ref{tbl:forcefields} have the following definitions: $b$ is the bond distance, $\theta$ is the bond angle, $\phi$ is the dihedral angle, $\delta$ is the multiplicity factor, and $\psi$ is the supplementary angle of $\phi$. $k_{b}$, $k_{\theta}$, $k_{\phi}$ are the force constants of the bond stretching, angle vibration, and dihedral potentials. $b_0$ and $\theta_0$ represent the equilibrium bond distance and bond angle, respectively. $C_{n}$ with $n$ ranging from 0 to 5 represents the coefficients for the Ryckaert-Bellemans dihedral potential~\cite{ryckaert1978}. $q$ represents the atomic partial charges of the electrostatic energy~($E_\text{{electrostatic}}$) term. A Lennard-Jones~(L-J) potential is used for the long-range van der Waals interactions, in which $\sigma$ represents the distance at which the particle-particle interaction energy is zero, and $\epsilon$ represents the depth of the potential well. The mixing rules for the L-J parameters for two dissimilar non-bonded atoms are given by Lorentz-Berthelot~\cite{lorentz_ueber_1881}~[ $\sigma_{ij} = \frac{\sigma_{i}+\sigma_{j}}{2}$, $\epsilon_{ij} = \sqrt{\epsilon_{i}\epsilon_{j}}$] for the force field by Orabi $\&$ English and geometric average~[$\sigma_{ij} = \sqrt{\sigma_{i}\sigma_{j}}$, $\epsilon_{ij} = \sqrt{\epsilon_{i}\epsilon_{j}}$] for the force field by Cordeiro. The values of these parameters are provided~(using the GROMACS convention) in Tables~S1 and S2 of the Supporting Information~(SI) for both force fields. The cutoff radius for Lennard-Jones and Coulombic interactions was set to 9~\AA. The Particle-Mesh-Ewald~\cite{darden1993particle, essmann1995smooth} method was used to treat long-range electrostatic interactions. Long-range tail corrections were applied to both energies and pressures~\cite{frenkel2023understanding}.

We use three different rigid water force fields in this study, namely TIP3P,~\cite{jorgensen1983comparison,neria1996simulation}TIP4P/2005,~\cite{abascal2005general} and TIP5P-E~\cite{rick2004reoptimization,mahoney2000five}. We also use a modified TIP3P water force field~(mTIP3P)~\cite{neria1996simulation} which was used in the work by Orabi $\&$ English~\cite{orabi_simple_2018}. In the remainder of this manuscript, the force field developed by Cordeiro~\cite{cordeiro_reactive_2014} is referred to as ``Cordeiro'' and the force field developed by Orabi $\&$ English~\cite{orabi_simple_2018} is referred to as ``Orabi''. 

\subsection{MD simulations}
\begin{figure}
    \centering
    \includegraphics[scale=1]{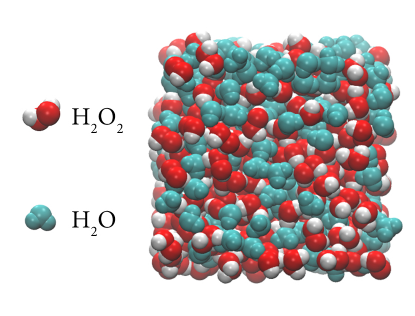}
    \caption{A snapshot of a simulation box containing 250 H\textsubscript{2}O\textsubscript{2} (red and white spheres represent oxygen and hydrogen atoms) and 250 H\textsubscript{2}O molecules (green spheres), generated by the Visual Molecular Dynamic~(VMD)
software~\cite{humphrey1996vmd}.}
    \label{fig:snapshot}
\end{figure}

All-atom Molecular Dynamics~(MD) simulation of anhydrous H\textsubscript{2}O\textsubscript{2} for a range of temperatures from 253~K to 353~K, and H\textsubscript{2}O\textsubscript{2} aqueous solutions for various mole fractions of H$_2$O$_2$ in the range from 0 to 1.0 were performed using the GROningen MAchine for Chemical Simulations~(GROMACS) version 2022.4~\cite{ berendsen_gromacs_1995, lindahl2001gromacs, van2005gromacs, hess2008gromacs, pronk_gromacs_2013}. Each system was prepared in a simulation box with an initial length of 27.6~\AA,~containing 500 molecules. A snapshot of a simulation box containing 250 H$_2$O$_2$ molecules and 250 H$_2$O molecules is shown in Figure~\ref{fig:snapshot}.  %

After energy minimization using the steepest descent algorithm followed by a conjugate gradient algorithm, the MD simulations were run for 100~ps in the constant number of atoms$\slash$molecules, volume and temperature~(\text{NVT}) ensemble. The simulations were then continued in the constant number of atoms$\slash$molecules, pressure and temperature~(\text{NPT}) ensemble for 25~ns. For calculating the viscosities and self-diffusivities, the simulations were continued in the NVT ensemble for another 20~ns.
The temperature was kept fixed by Nos\'{e}-Hoover thermostat~\cite{hoover_canonical_1985}. 
The Parinello-Rahman barostat~\cite{parrinello_polymorphic_1981} with a time constant of 1~ps and compressibility of 4.5~$\times$~10$^{-5}$~bar$^{-1}$ was used to keep the pressure at 1~bar.
In all simulations, the Newton's equations of motion were integrated with a leap-frog~\cite{eastwood_computer_2021} algorithm with a time step of 2~fs. 
Periodic boundary conditions were applied in all Cartesian directions. 
The parallel linear constraint solver~(P-LINCS)~\cite{hess1997lincs, hess2008p} was used to constrain H bonds. 

\subsection{MC simulations}
Continuous Fractional Component Monte Carlo~(CFCMC) simulations~\cite{shi_continuous_2007, shi_improvement_2008,rahbari_recent_2021} using the open-source Brick-CFCMC software~\cite{rahbari_recent_2021, hens_brick-cfcmc_2020, polat_new_2021} were performed in the isothermal-isobaric~(NPT) ensemble. In the CFCMC technique, fractional molecules~(compared to normal or ``whole'' molecules) are introduced whose interactions with the rest of the system are scaled with a continuous coupling parameter~$\lambda$~($\lambda \in [0,1]$). 
 The minimum value of $\lambda$ ($\lambda$ = 0) indicates no interactions between the fractional molecule and the rest of the molecules in the system~(i.e., fractional molecules act as ideal gas molecules). $\lambda$ = 1 represents full interactions between the fractional molecules and the other molecules in the system~(i.e., the fractional molecule acts as whole molecules). The coupling parameter $\lambda$ is biased with a weight function~($W(\lambda)$) using the Wang-Landau algorithm~\cite{wang_efficient_2001} to improve molecule transfers~(insertions$\slash$deletions). 
 This ensures a smooth observed probability distribution of $\lambda$. 
 We used 100 bins to construct a histogram for the $\lambda$ values and its probability of occurrence (p($\lambda$)).
 The Boltzmann average of any property~(A) is then computed using~\cite{poursaeidesfahani_direct_2016}
\begin{equation}
    \langle A \rangle = \frac{\langle~A~\text{exp}[-W(\lambda)]\rangle_{\text{biased}}}{\langle\text{exp}[-W(\lambda)]\rangle_{\text{biased}}}.
\end{equation}
The chemical potential of species $i$ is calculated with respect to its ideal gas chemical potential~\cite{hens_brick-cfcmc_2020}
\begin{equation}
    \mu_{i} = \mu_{i}^{\text{ideal}} + \mu_{i}^{\text{ex}},
\end{equation}
where $\mu_{i}^{\text{ideal}}$ and $\mu_{i}^{\text{ex}}$ are the ideal gas and excess chemical potential of the species $i$, respectively.  
The excess chemical potential can be related to the Boltzmann sampled probability distribution of $\lambda$ by the following equation~\cite{hens_brick-cfcmc_2020}
\begin{equation}
    \mu_{i}^{\text{ex}} = -k_{\text{B}}T~\text{ln}~\frac{p(\lambda=1)}{p(\lambda=0)},
\end{equation}
where $p(\lambda$=1) and $p(\lambda$=0) are the Boltzmann sampled probability distributions of $\lambda$ at 1 and 0, respectively. $k_{\text{B}}$ is the Boltzmann constant, and \textit{T} is the absolute temperature. The excess chemical potential at infinite dilution ($\mu^{\text{ex},\infty}$) can be used to determine the Henry volatility coefficient\footnote{The definition of the Henry law constants is according to Sander~\cite{sander2015compilation}.} ($K^{\text{px}}_{v}$) by\cite{salehi_computation_2020}

\begin{equation}
    K^{\text{px}}_{v}  = \rho~k_{\text{B}}T\exp{\left( \frac{\mu^{\text{ex},\infty}}{k_{\text{B}}T}\right)},
\end{equation}
where $\rho$ is the number density of the solvent. This yields the Henry volatility coefficient ($K^{\text{px}}_{v}$) in units of \text{[Pa]}. The Henry coefficient~($H^{\text{cp}}_{s}$) in units of [mol/m\textsuperscript{3}Pa] can be obtained using the following conversion: $H^{\text{cp}}_{s}\approx \frac{\rho_{\text{H}_2\text{O}}}{M_{\text{H}_2\text{O}} K^{\text{px}}_{v}}$ in which $\rho_{\text{H}_2\text{O}}$ is the density of water, and $M_{\text{H}_2\text{O}}$ is the molar mass of water~\cite{sander2015compilation}. 


CFCMC simulations contained 300 water molecules 
in a cubic simulation box with initial dimensions of 21~\AA. A single fractional molecule of H\textsubscript{2}O\textsubscript{2} was introduced to calculate its excess chemical potential. The cut-off radius for the intermolecular L-J and Coulombic interactions was set to 9~\AA. 
The Ewald summation~\cite{wells2015ewald} method was used for calculating electrostatic interactions. 
Long-range tail corrections were applied to the L-J potential. 
Periodic boundary conditions were applied in all directions. 

For CFCMC simulations, 1,000 initialization cycles were carried out followed by $5\times10^{6}$ equilibration cycles and $5\times10^{6}$ production cycles. One cycle refers to \textit{N} number of trial moves, where \textit{N} is the total number of molecules. Trial moves were selected with the following probabilities: 
32$\%$ translation moves, 22$\%$ rotation moves, 1$\%$ volume changes, 5$\%$ each of bending and torsion moves, 25$\%$~$\lambda$ changes and 10$\%$ hybrid moves that combined swap and identity change moves~\cite{hens_brick-cfcmc_2020}. Three independent simulations were performed for each combination of water force field and  H\textsubscript{2}O\textsubscript{2} force field to obtain an average value and the standard deviation for the Henry coefficients.
\section{Results and Discussion}
\label{FF validation}


\subsection{Densities} \label{sec:Density} 
\begin{figure}
    \centering
    \includegraphics[scale=0.55]{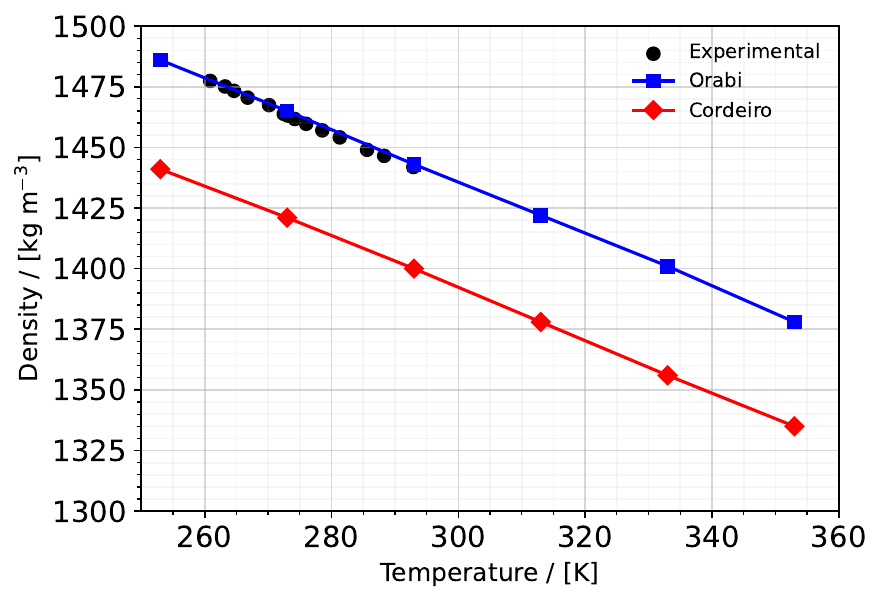}
    \caption{Densities of anhydrous H$_2$O$_2$ at various temperatures using the Cordeiro~\cite{cordeiro_reactive_2014} and Orabi~\cite{orabi_simple_2018} force fields at 1~bar with the experimental values~\cite{maass1920properties}. Error bars are estimated based on the standard deviation and are much smaller than the markers used in the figure.}
    \label{fig:DvT_AnHP}
\end{figure}

The densities of anhydrous H\textsubscript{2}O\textsubscript{2} for a temperature range of 253~K to 353~K~(in steps of 20~K) for both the Orabi and Cordeiro force fields are plotted in  Figure~\ref{fig:DvT_AnHP}. We used the \textit{gmx density} tool to compute the average density of each system. The experimental values are shown in black circles~\cite{maass1920properties}. The melting point and boiling point of  H\textsubscript{2}O\textsubscript{2} are reported as 272.74~K and 423.15~K, respectively~\cite{giguere1983molecular}.  
While the Cordeiro force field underestimates the densities of anhydrous H\textsubscript{2}O\textsubscript{2} by about 3$\%$ compared to the experimental values, the densities of anhydrous H\textsubscript{2}O\textsubscript{2} using the Orabi force field are in excellent agreement with the experimental values. 


\begin{figure}
    \centering
    \includegraphics[scale=1.6]{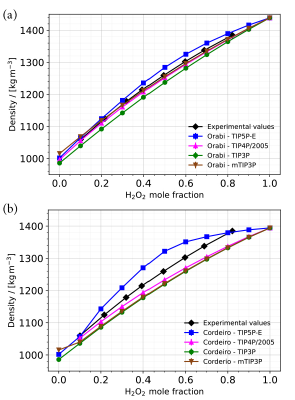}
    \caption{Densities of H\textsubscript{2}O\textsubscript{2} aqueous solution for various mole fractions of H\textsubscript{2}O\textsubscript{2} at $T=298$~K and 1~bar using the 
    (a) Orabi~\cite{orabi_simple_2018} and (b) Cordeiro~\cite{cordeiro_reactive_2014} force fields in combination with the TIP3P~\cite{jorgensen1983comparison}, mTIP3P~\cite{neria1996simulation}, TIP4P/2005~\cite{abascal2005general} and TIP5P-E~\cite{rick2004reoptimization,mahoney2000five} water force fields. Experimental values~\cite{easton_behaviour_1952} are added for comparison. Error bars are estimated based on the standard deviation and are much smaller than the markers used in the figure.}
    \label{fig:Dvmf_W_HP}
\end{figure}

Next, we evaluate which water force field is suitable for predicting the densities of H\textsubscript{2}O\textsubscript{2} aqueous solutions. To this end, we modelled systems of H\textsubscript{2}O\textsubscript{2} aqueous solutions for the whole range of H\textsubscript{2}O\textsubscript{2} mole fractions~(0 to 1.0) at $T=298$~K and 1~bar using the Orabi or Cordeiro force fields in combination with four different water force fields: TIP3P, TIP4P/2005, TIP5P-E, and the modified TIP3P water force field~(mTIP3P) which was used in the work by Orabi $\&$ English~\cite{orabi_simple_2018}. 
The choice of temperature at 298~K and pressure at 1~bar was motivated by the availability of experimental results by which we could validate our models. The results as a function of the H$_2$O$_2$ mole fraction are shown in Figure~\ref{fig:Dvmf_W_HP}. 
The experimental values~\cite{easton_behaviour_1952} are added for comparison. 
 
The densities of pure water~(i.e., a mole fraction of zero), using the four water force fields are in good agreement with the reported data at 298~K and 1~bar~\cite{abascal2005general, vega2005relation}.
The Orabi force field for H\textsubscript{2}O\textsubscript{2} in combination with the TIP4P/2005 water force field or mTIP3P water force field predicts the densities of the aqueous solutions in good agreement (ca.~$0.6\%$) with the experimental values\cite{easton_behaviour_1952}. 
The predicted values for densities of solutions using the TIP5P-E water force field in combination with the Orabi or Cordeiro force fields at low and high concentrations of H\textsubscript{2}O\textsubscript{2} 
are in good agreement with the experimental values. At intermediate concentrations~(0.4~-~0.6 mole fractions), however, the TIP5P-E in combination with the Orabi or Cordeiro force fields overestimates the densities of solutions by 2$\%$ and 5$\%$, respectively. 
The TIP3P water force field in combination with the Orabi force field underestimates the densities of the solutions by 2$\%$. The Cordeiro-TIP3P and Cordeiro-mTIP3P models underestimate the densities by 3$\%$ at intermediate concentrations. The Cordeiro force field in combination with the TIP4P/2005 water force field also underestimates the densities of the solution with a more pronounced effect at higher mole fractions of H$_2$O$_2$~($\ge 0.5$, by 3$\%$). 

We conclude that the Orabi force field is a better force field than the Cordeiro force field for predicting the densities of pure H$_2$O$_2$ in the temperature range of 253~-~353~K. In addition, the TIP4P/2005 or the mTIP3P force field in combination with the Orabi force field predicts the densities of H$_2$O$_2$ aqueous solutions for the whole range of mole fractions~(0~-~1.0) in very good agreement with the experimental values. 



\subsection{Viscosities}
\label{sec:viscosity}

\begin{figure*}
    \centering
    \includegraphics[scale=1.7]{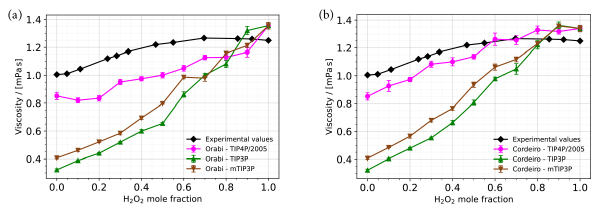}
    \caption{Viscosities of H\textsubscript{2}O\textsubscript{2} aqueous solution for various mole fractions of H$_2$O$_2$ at 293~K for 
    the (a) Orabi~\cite{orabi_simple_2018} and (b) Cordeiro~\cite{cordeiro_reactive_2014} force fields in combination with the TIP3P~\cite{jorgensen1983comparison}, mTIP3P~\cite{neria1996simulation} and TIP4P/2005~\cite{abascal2005general}. The results including the TIP5P-E water force field are shown in Figure S1 of the Supplementary Information. 
    Error bars are estimated based on the standard deviation. The experimental values~\cite{phibbs1951hydrogen} at 293~K are added for comparison.}
    \label{fig:Vvmf_NTIP5P_W_HP}
\end{figure*}

We used the \textit{gmx energy} tool to compute the viscosities~\cite{hess2002determining}. This tool uses the Navier-Stokes equation in which an external force is applied to the system. This causes a velocity gradient in the system from which the viscosity can be calculated~\cite{hess2002determining}.  
The viscosities of H\textsubscript{2}O\textsubscript{2} aqueous solutions for various H$_2$O$_2$ mole fractions~(0 to 1.0) at $T=$293~K were computed.  
Figure \ref{fig:Vvmf_NTIP5P_W_HP} shows the values of viscosity using the Orabi (a) and Cordeiro (b) force fields in combination with the TIP3P, mTIP3P and TIP4P/2005 force fields. The results including the TIP5P-E water force field are shown in Figure S1 of the Supplementary Information.  
The standard deviation is used to estimate error bars. The experimental values at 293~K are included for comparison~\cite{phibbs1951hydrogen}. 

The viscosities of pure water~(i.e., mole fraction = 0 ) are in good agreement with the computed values using the TIP3P, mTIP3P, TIP4P/2005, and TIP5P-E force fields~\cite{gonzalez2010shear}. 
The experimental value of the viscosity of pure H$_2$O$_2$ at 293~K is 1.25~mPa\,s~\cite{phibbs1951hydrogen}. The computed value is 1.36~mPa\,s by using the Orabi force field, and is 1.34~mPa\,s by using the Cordeiro force field. 

The combination of the Orabi force field with the mTIP3P or TIP4P/2005 underestimates the viscosities of H\textsubscript{2}O\textsubscript{2} aqueous solutions for mole fractions up to 0.9. The Orabi-TIP3P model underestimates the values up to a mole fraction of 0.8, above which it slightly overestimates. The combination of Cordeiro force field with the mTIP3P or TIP3P or TIP4P/2005 water force fields follows a similar trend. The Cordeiro-mTIP3P and Cordeiro-TIP3P models underestimate the viscosities up to a mole fraction of 0.8, above which it slightly overestimates. The Cordeiro-TIP4P/2005 model, however, underestimates the values of viscosities by 7$\%$ up to a mole fraction of 0.5 while it overestimates by ca.~5$\%$ for H\textsubscript{2}O\textsubscript{2} mole fractions higher than 0.8.
Contrary to the other water force fields, the TIP5P-E water force field in combination with either the Orabi or Cordeiro force fields predicts a relatively high peak in viscosity at the intermediate mole fractions~(mole fraction of 0.5), see Figure S1 of the Supplementary Information. This may be due to structural changes which the TIP5P-E water force field induces in the system. This is addressed in section~\ref{sec:RDFs} using radial distribution functions.

We conclude that the TIP4P/2005 water force field in combination with the Orabi force field or the Cordeiro force field predicts the viscosities of H$_2$O$_2$ aqueous solutions in better agreement with the experimental values.
\subsection{Self-diffusion coefficients} Diffusion coefficients were calculated from the mean-squared displacements (MSD), and were corrected for finite-size effects with the Yeh-Hummer equation~\cite{yeh2004system,celebi2021finite}
\begin{equation}
    D = D_{\text{MD}} + \frac{k_{\text{B}}T\xi}{6\pi\eta L},
\end{equation}
where $D$ and $D_{\text{MD}}$ denote the diffusion coefficient calculated with and without the finite-size effects corrections, respectively. $k_{\text{B}}$ is the Boltzmann constant, $T$ is the absolute temperature~(in K), $\xi$ is a dimensionless number which for a cubic simulation box is equal to 2.837, $L$ is the length of the cubic simulation box, and $\eta$ is the viscosity of the system. We used the \textit{gmx msd} tool to obtain the MSD as a function of time. $D_{\text{MD}}$ is obtained by fitting the MSD to
    \begin{equation}
        \langle r^2 \rangle_{\text{MSD}}= 2dD_{\text{MD}}t,
        \label{equ:diffusion}
\end{equation}
where $d=3$ is the dimension of the system. The self-diffusion coefficients were calculated from 1~ns to 20~ns NVT trajectories. Figure S2 of the SI shows an example of MSD versus time on a logarithmic scale.
 \begin{figure}
    \centering
    \includegraphics[scale=1.4]{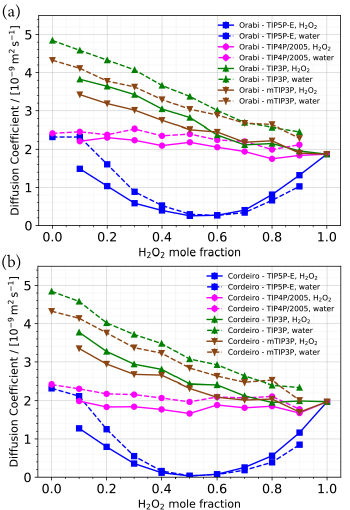}
    \caption{Self-diffusion coefficients of water molecules and H\textsubscript{2}O\textsubscript{2} molecules in H\textsubscript{2}O\textsubscript{2} aqueous solutions for different mole fractions of H$_2$O$_2$ at 298~K using the (a) Orabi~\cite{orabi_simple_2018} and (b) Cordeiro~\cite{cordeiro_reactive_2014} force fields in combination with the TIP3P~\cite{jorgensen1983comparison}, mTIP3P~\cite{neria1996simulation}, TIP4P/2005~\cite{abascal2005general} and TIP5P-E~\cite{rick2004reoptimization,mahoney2000five} water force fields. Error bars are estimated based on the standard deviation
    and are much smaller than the markers used in the figure.}
    \label{fig:DCvmf_W_HP}
\end{figure}
Figure \ref{fig:DCvmf_W_HP} shows the self-diffusion coefficients of H\textsubscript{2}O\textsubscript{2} and water in aqueous H\textsubscript{2}O\textsubscript{2} solutions for the whole range of hydrogen peroxide mole fractions~(0 to 1.0).  
The self-diffusion coefficients of pure water~(i.e., mole fraction=0) for the four water force fields are in good agreement with the values reported in Ref.~\cite{abascal2005general, mark2001structure, rick2004reoptimization}. The TIP5P-E and TIP4P/2005 water force fields predict the value of the self-diffusion coefficient in better agreement with the experimental value~(2.3$\times 10^{-9}$m\textsuperscript{2}/s~\cite{holz2000temperature}). 

The self-diffusion coefficients of both the water and the H$_2$O$_2$ molecules decrease monotonically by increasing the mole fraction of H$_2$O$_2$ using the Orabi-TIP3P or Orabi-mTIP3P models. There is a similar trend for the Cordeiro-TIP3P and Cordeiro-mTIP3P models. The TIP4P/2005 water force field in combination with either the Orabi force field or the Cordeiro force field predicts a relatively constant self-diffusion coefficient for both water and H$_2$O$_2$ for the whole range of mole fractions. This is in agreement with a recent experimental study~\cite{suresh2022production} where it was concluded that the self-diffusion coefficients of H\textsubscript{2}O\textsubscript{2} in solutions are insensitive to its concentration.
The TIP5P-E water force field in combination with either the Orabi or Cordeiro force fields predicts a minimum at a mole fraction of 0.5 for the self-diffusion coefficients of both water and H$_2$O$_2$. This is correlated with its very high value of viscosities~(see Figure S1 of the Supplementary Information).

\subsection{Radial Distribution Functions}
\label{sec:RDFs}
\begin{figure}[!htb]
    \centering
     \includegraphics[width=0.48\textwidth]{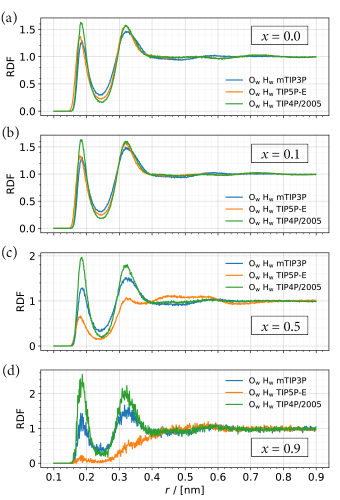}
    \caption{Radial distribution functions~(RDFs) as a function of radial distance, $r$~[nm], for O$_{\text w}$~(O of water)~-~H$_{\text w}$~(H of water) for H\textsubscript{2}O\textsubscript{2} aqueous solutions with $x=$~0.1~(b), $x=$~0.5~(c), and $x=$~0.9~(d) at 298~K and 1 bar using the Orabi force field in combination with the mTIP3P~\cite{neria1996simulation}, TIP5P-E~\cite{rick2004reoptimization,mahoney2000five}, and TIP4P/2005~\cite{abascal2005general} water force fields, where $x$ is the mole fraction of H\textsubscript{2}O\textsubscript{2}. The RDF for pure water is plotted in (a).}
    \label{fig:RDFWater}
\end{figure}

\begin{figure}[!htb]
    \centering
     \includegraphics[width=0.48\textwidth]{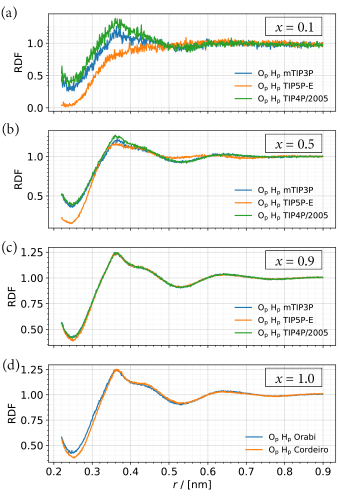}
    \caption{Radial distribution functions~(RDFs) as a function of radial distance, $r$~[nm], for O$_{\text p}$~(O of H\textsubscript{2}O\textsubscript{2})~-~H$_{\text p}$~(H of H\textsubscript{2}O\textsubscript{2}) for H\textsubscript{2}O\textsubscript{2} aqueous solutions with $x=$~0.1~(b), $x=$~0.5~(c), and $x=$~0.9~(d) at 298~K and 1 bar using the Orabi~\cite{orabi_simple_2018} force field in combination with the mTIP3P~\cite{neria1996simulation}, TIP5P-E~\cite{rick2004reoptimization,mahoney2000five}, and TIP4P/2005~\cite{abascal2005general} water force fields, where $x$ is the mole fraction of H\textsubscript{2}O\textsubscript{2}. The first peak (at ca. 0.19~nm) was removed to distinguish the differences between the combinations clearly. The RDF for pure H$_2$O$_2$ is plotted in (d) using the Orabi force field and the Cordeiro force field.}
    \label{fig:RDFHP}
\end{figure}

\begin{figure*}[!ht]
    \centering
   \includegraphics[width=0.95\textwidth]{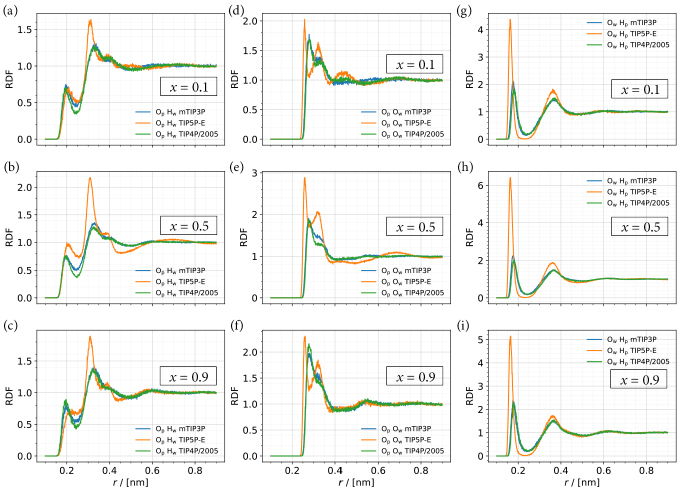}
    \caption{Radial distribution functions~(RDFs) as a function of radial distance, $r$~[nm], for O$_{\text p}$~(O of H$_2$O$_2$) and H$_{\text w}$~(H of water) (a~-~c), O$_{\text w}$~(O of water) and O$_{\text p}$~(O of H$_2$O$_2$) (d~-~f), and O$_{\text w}$~(O of water) and H$_{\text p}$~(H of H$_2$O$_2$) (g~-~i) for systems using the mTIP3P~\cite{neria1996simulation}, TIP5P-E~\cite{rick2004reoptimization,mahoney2000five}, and TIP4P/2005~\cite{abascal2005general} water force fields in combination with the Orabi force field for $x=$~0.1, $x=$~0.5, and $x=$~0.9 at 298~K and 1 bar, where $x$ is the mole fraction of H\textsubscript{2}O\textsubscript{2}.}
    \label{fig:RDFCollection}
\end{figure*}
Structural properties of H$_2$O$_2$ aqueous solution at various mole fractions were investigated using the radial distribution functions~(RDF). Note that there are 4 atom types in each system: hydrogen of water~(H$_{\text w}$), oxygen of water (O$_{\text w}$), hydrogen of H$_2$O$_2$~(H$_{\text p}$), and oxygen of H$_2$O$_2$~(O$_{\text p}$).

The RDFs for O$_{\text w}$ and H$_{\text w}$ in pure water using the mTIP3P, TIP5P-E and TIP4P/2005 water force fields are shown in Figure~\ref{fig:RDFWater} (a). The results show a first peak at approximately 0.18~nm, and a second peak at approximately 0.32~nm. The peak heights slightly differ between the water force fields with the TIP4P/2005 predicting a higher value followed by the TIP5P-E, and then the mTIP3P predicting a smaller value. 

The respective RDFs in H\textsubscript{2}O\textsubscript{2} aqueous solutions using the Orabi force field in combination with the three water force fields for systems with mole fractions of 0.1, 0.5 and 0.9 are also shown in Figure~\ref{fig:RDFWater} (b, c, d). The RDFs for O$_{\text w}$ and H$_{\text w}$ for systems using the Cordeiro force field in combination with the mTIP3P, TIP5P-E and TIP4P/2005 are shown in Figure S3 of the SI. In the system with a mole fraction of 0.1, the position of the peaks is independent of the water force fields. As the mole fraction of H$_2$O$_2$ is increased to 0.5 and 0.9, the position of the peaks does not change in systems where the mTIP3P or TIP4P/2005 force field is used. In the systems where the TIP5P-E water force field is used, however, the RDF changes: the height of the peaks become smaller and an additional structural correlation appears in the system with a mole fraction of 0.5. This additional correlation between the water molecules persists till 0.8~nm, whereas for the systems in which the mTIP3P or TIP4P/2005 is used, the structural correlation persists only till 0.6~nm.  In the Orabi-TIP5P-E system with a mole fraction 0.9, the first two peaks disappear. A similar trend can be observed for the combinations involving the Cordeiro force field with the mTIP3P, TIP5P and TIP4P/2005 water force fields (see Figure~S3 of SI). In the Cordeiro-TIP5P-E combination with a mole-fraction of 0.5, however, the structural correlation between the water molecules is stronger than that of the corresponding Orabi-TIP5P-E combination. This can be seen from a more prominent peak after 0.4~nm compared to the other models. 

The RDFs for O$_{\text p}$ and H$_{\text p}$ in H$_2$O$_2$ aqueous solution with mole fractions of 0.1, 0.5 and 0.9 using the Orabi force field in combination with the three water force fields are shown in Figure~\ref{fig:RDFHP} (a, b, c). Similarly, the respective RDFs using the Cordeiro force field in combination with the three water force fields are shown in Figure~S4 of SI. The RDF of O$_{\text p}$ and H$_{\text p}$ in pure H$_2$O$_2$ using the Orabi and the Cordeiro force fields are also shown in Figure~\ref{fig:RDFHP} (d). The first peak in the RDF has a large amplitude, therefore we removed it to be able to distinguish the differences between the systems more clearly~(see Figure S5 of SI). RDFs of the system at a mole fraction of 0.9 are almost identical using the three different water force fields with a second peak at 0.37~nm. By decreasing the mole fraction of H$_2$O$_2$ to 0.5, and 0.1, the RDFs remain the same for the systems in which the mTIP3P-E or the TIP4P/2005 is used. For the system in which the TIP5P-E water force field is used, however, the RDF changes drastically. This is also the case with the Cordeiro-TIP5P-E model. The RDF for pure H$_2$O$_2$ using the Orabi force field is almost identical as the system using the Cordeiro force field.  

The RDFs for O$_{\text p}$ and H$_{\text w}$, O$_{\text p}$ and O$_{\text w}$, and  O$_{\text w}$ and H$_{\text p}$ using the Orabi force field with the three water force fields~(mTIP3P, TIP4P/2005, and TIP5P-E) for solutions with mole fractions of 0.1, 0.5 and 0.9, are shown in Figure~\ref{fig:RDFCollection}. Likewise, RDFs with the Cordeiro force field and the water force fields are shown in Figure~S6 of SI. In systems where the mTIP3P and TIP4P/2005 water force fields were used, RDFs have the same structure in which the position of the first peak is in good agreement with X-ray measurements on crystals of H\textsubscript{2}O\textsubscript{2}$\cdot$2H\textsubscript{2}O~\cite{olovsson1960crystal} and simulation results~\cite{orabi_simple_2018}. On the contrary, the structural properties in systems where the TIP5P-E force field was used, have changed. A comparable effect can be seen in Figure~S6 of SI where the Cordeiro-TIP5P-E combination is used. 
The number of water molecules in the micro and first solvation shells of H\textsubscript{2}O\textsubscript{2} molecule were obtained by integrating up to the first and second minima of the RDF for O\textsubscript{p}~-~O\textsubscript{w}, respectively. The results are shown in Table S3 and S4 of the SI. Orabi~$\&$~English~\cite{orabi_simple_2018} reported 3.0 and 9.4 water molecules in the micro and first solvation shells, respectively, for a single peroxide molecule in 500 water molecules using the mTIP3P water force field. Authors in Ref.~\cite{moin2012ab} report 6.0 water molecules in the first solvation shell of  H\textsubscript{2}O\textsubscript{2} using a hybrid quantum-classical simulation. 
According to our results, the number of water molecules in the first solvation shell at mole fractions of 0.1 and 0.9 are lower for systems where the TIP5P-E water force field is used. At a mole fraction of 0.5, however, the number of water molecules in the first solvation shell slightly increases. 

Our results suggest that the addition of the TIP5P-E water force field to the H$_2$O$_2$ Orabi or Cordeiro models disturbs the structural properties of the systems such that there is a 
stronger interaction between the water molecules and H$_2$O$_2$. This effect is more pronounced at a mole fraction of 0.5 which is correlated with the prediction of the Orabi-TIP5P-E and Cordeiro-TIP5P-E models for densities, viscosities, and self-diffusion coefficients~(see Figure~\ref{fig:Dvmf_W_HP}, \ref{fig:Vvmf_NTIP5P_W_HP} and \ref{fig:DCvmf_W_HP}). 

\subsection{Hydrogen Bond analysis}
\begin{figure}
    \centering
    \includegraphics[scale=1.4]{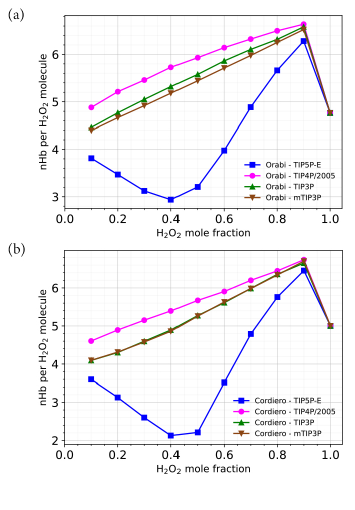}
    \caption{Number of hydrogen bonds per H\textsubscript{2}O\textsubscript{2} molecule for systems with various mole fractions of H\textsubscript{2}O\textsubscript{2} at $T=298$~K and 1~bar using the 
    (a) Orabi~\cite{orabi_simple_2018} and (b) Cordeiro~\cite{cordeiro_reactive_2014} force fields in combination with the TIP3P~\cite{jorgensen1983comparison}, mTIP3P~\cite{neria1996simulation}, TIP4P/2005~\cite{abascal2005general} and TIP5P-E~\cite{rick2004reoptimization,mahoney2000five} water force fields.}
    \label{fig:Hbonds_mf_HP}
\end{figure}
The number of hydrogen bonds (calculated as the summation of hydrogen bonds between H\textsubscript{2}O\textsubscript{2}~-~H\textsubscript{2}O\textsubscript{2}, H\textsubscript{2}O\textsubscript{2}~-~water and water~-~water) per H\textsubscript{2}O\textsubscript{2} molecule  for combinations of the Orabi and Cordeiro force fields with the water force fields is shown in Figure~\ref{fig:Hbonds_mf_HP}. We used the geometric criterion for hydrogen bonds proposed in Ref.~\cite{luzar1993structure}. The number of hydrogen bonds for both the Orabi and Cordeiro force fields in combination with the TIP4P/2005, TIP3P and mTIP3P water force fields exhibit a steady increase until a mole fraction of 0.9.  Existing literature indicates the existence of about 4 hydrogen bonds between hydrogen peroxide and water molecules~\cite{moin2012ab, priyadarsini_structure_2022}. For pure H\textsubscript{2}O\textsubscript{2} systems, the number of hydrogen bonds sharply decreases to about 5 for both the Orabi and Cordeiro systems. The Orabi-TIP5P-E and Cordeiro-TIP5P-E combinations are however different compared to the others. These combinations indicate a minimum in the number of hydrogen bonds at a mole fraction of 0.4 (around 3 hydrogen bonds for Orabi-TIP5P-E and 2 for Cordeiro-TIP5P-E). This minima for the number of hydrogen bonds at the intermediate concentrations for both these systems are concurrent with the high viscosities and low diffusion seen earlier.
Analysis of the RDFs, solvation shells and the number of hydrogen bonds of the TIP5P-E systems indicate a variation in the arrangement of water molecules around a H\textsubscript{2}O\textsubscript{2} molecule. The effects of this structural difference is reflected in the incongruity of the viscosities and self-diffusion coefficients of the TIP5P-E systems with respect to other water force fields based systems. 


\subsection{Henry coefficients}



\begin{table}
\begin{center}

  \caption{Excess chemical potentials~($\mu^{\text{ex}}$), the Henry volatility coefficient~($K^{\text{px}}_{v}$), and the Henry coefficient~($H^{\text{cp}}_{s}$), using the Orabi and the Cordeiro force fields in combination with the TIP4P/2005,TIP3P and mTIP3P water force fields. Errors are estimated using standard deviations of independent simulations. }
 \label{tbl:Solubility}
\footnotesize
  \begin{tabular}{@{\extracolsep{\fill}}cccc}
    \hline
	Model  & $\mu^{\text{ex}}$/[K] & $K^{\text{px}}_{v}$/[Pa] & $H^{\text{cp}}_{s}$/[mol/m\textsuperscript{3}\,Pa]  \\
		\hline
		Orabi~-~TIP4P/2005 &$-3734\pm30$ & $512\pm52$ & $109\pm11$\\
  		Orabi~-~TIP3P & $-3836\pm27$ & $365\pm32$ & $153\pm14$  \\
		Orabi~-~mTIP3P & $-3963\pm14$ & $239\pm11$ & $232\pm11$
    \vspace{3mm}\\
          Cordeiro~-~TIP4P/2005 &$-4142\pm32$ & $132\pm14$ & $424\pm46$ \\
		Cordeiro~-~TIP3P & $-4392\pm58$ & $58\pm11$ & $989\pm190$  \\
		Cordeiro~-~mTIP3P & $-4487\pm57$ & $42\pm7$ & $1357\pm275$\\
  \hline
  \end{tabular}
  \end{center}
\end{table}

The Henry coefficients were computed for H$_2$O$_2$ in water using the Orabi and Cordeiro force fields in combination with the various water force fields. It should be noted that we have not further considered the TIP5P-E water force field for the solubility calculations as it has been ascertained from the earlier sections that neither the Cordeiro nor Orabi force fields in combination with the TIP5P-E water force field could accurately predict the densities, viscosities and self-diffusion coefficients of H\textsubscript{2}O\textsubscript{2} aqueous systems. The results of the solubility calculations are provided in Table~\ref{tbl:Solubility}. The reported values range from 670 to 1400~[mol/m$^{\text{3}}$\,Pa]~\cite{hwang1985thermodynamics, yoshizumi1984measurements, zhou1992aqueous, huang2010reinvestigation}. It is evident that the Cordeiro force field in combination with the TIP3P or mTIP3P water force fields predicts the Henry constants that are within the range of the reported experimental values. 
\section{Conclusions}
\label{Conclusion}
We performed MD and CFCMC simulations to study thermodynamics properties of aqueous solutions of H$_2$O$_2$. The quality of the available force fields of H$_2$O$_2$, Cordeiro~\cite{cordeiro_reactive_2014} and Orabi~\cite{orabi_simple_2018}, was evaluated by comparing the results with experiments. The densities of pure H$_2$O$_2$ computed using the Orabi force field are in excellent agreement with the experimental values for the temperature range of 253~K to 353~K. The Cordeiro force field underestimates the densities of pure H$_2$O$_2$ by 3$\%$. We computed densities, viscosities, and self-diffusion coefficients of H$_2$O$_2$ in aqueous solutions for the whole range of mole fractions of H$_2$O$_2$~(0 to 1.0) at ambient temperatures and pressures using four water force fields: TIP3P, mTIP3P, TIP4P/2005, and TIP5P-E. The results show that the TIP4P/2005 water force field in combination with the Orabi force field can predict the densities of H$_2$O$_2$ aqueous solution in excellent agreement with experimental values. Both the Orabi and Cordeiro force fields in combination with the TIP4P/2005 water force field predict the viscosities of H$_2$O$_2$ in reasonable agreement with experimental results. The TIP5P-E water force field leads to a very high value (maximum) for viscosity of H$_2$O$_2$ aqueous solutions at a mole fraction of 0.5, and thereby a very small value (minimum) for self-diffusion coefficient of H$_2$O$_2$ and water. The TIP4P/2005 force field in combination with either of the Orabi or Cordeiro force fields predicts a relatively constant diffusion coefficient for the whole range of H$_2$O$_2$ mole fractions that is in agreement with a recent experimental study~\cite{suresh2022production}. We studied the structural properties of H$_2$O$_2$ aqueous solutions using radial distribution functions. These results suggest that the use of the TIP5P-E water force field in combination with either the Orabi or Cordeiro force field, predict a stronger interaction between water molecules and H$_2$O$_2$ molecules. Hydrogen bond analysis indicates a steady increase in the number of hydrogen bonds per H\textsubscript{2}O\textsubscript{2} molecule with increasing solute concentration for H\textsubscript{2}O\textsubscript{2} aqueous solutions. The Cordeiro-TIP5P-E and Orabi-TIP5P-E systems exhibited a minimum at the intermediate solute concentrations. This is in line with the deviation in the dynamic properties (viscosities and self-diffusion coefficients) of these systems. 
Finally, we computed the Henry coefficients of H$_2$O$_2$ in water. The values using the Cordeiro force field in combination with either of the TIP3P or mTIP3P water force fields are within the range of experimental values. The quantitative data presented in this work can be used by macroscopic plasma fluid models to determine the uptake of H$_2$O$_2$ from the gas phase plasma by liquid~\cite{tian2014atmospheric} or to interpret and complement experimental findings~\cite{machala2018chemical}. 

\section*{Author Contributions}
TS carried out the simulations and data analysis. All authors provided critical feedback on the interpretation of data analysis. BB conceived and supervised the project. TS and BB wrote the manuscript in collaboration with all the authors.  

\section*{Supplementary Information}
Force field parameters for Cordeiro and Orabi \& English using the GROMACS convention are provided in Tables S1 and S2, respectively; Figure S1 shows viscosities of H$_2$O$_2$ aqueous solutions for various mole fractions of H$_2$O$_2$; Figure S2 shows an example of the MSD versus time on a logarithmic scale; Radial distribution functions are illustrated in Figures S3-S6; The number of water molecules in the micro solvation shell, and the first solvation shell of H$_2$O$_2$ in aqueous solution for various mole fractions of H$_2$O$_2$ is shown in Tables S3 and S4, respectively.    
\section*{Conflicts of interest}
There are no conflicts to declare.




\section*{Acknowledgements}
BB thanks the strategic alliance between TU/e, Utrecht University, and University Medical Center Utrecht, and TS thanks the Institute for Complex Molecular Systems for financial support.


\renewcommand\refname{References}
\bibliography{references,Behnaz-added, rsc} 
\bibliographystyle{rsc} 








\newpage

\renewcommand{\figurename}{Figure.}
\renewcommand{\thetable}{S\arabic{table}}  
\renewcommand{\thefigure}{S\arabic{figure}} 
\setcounter{figure}{0}

\onecolumngrid 

\begin{center}
    \textbf{\Huge{Supporting Information}}

    \vspace{1.0cm}

    for

    \vspace{1.0cm}

    \textbf{\Large{Computing solubility and thermodynamics properties of H$_2$O$_2$ in water}}
\end{center}

\newpage

\begin{table*} \label{Rodrigo FF}
\small
  \caption{\ Parameters of potential energy functions corresponding to bond-stretching, bond-bending, torsional, van der Waals and electrostatic energies developed by Cordeiro~\cite{cordeiro_reactive_2014} for H\textsubscript{2}O\textsubscript{2}.}
  \label{tbl:}
  \begin{tabular*}{\textwidth}{@{\extracolsep{\fill}}llll}
    \hline
    Parameter & Functional form & Units & Value\\
    \hline
         $b_{\text{OH}}$ &$\frac{1}{4}k_{b}(b\textsuperscript{2}-b\textsubscript{0}\textsuperscript{2})\textsuperscript{2}$& $[\text{nm}]$& 0.0981 \\
         $k_{\text{b(OH)}}$ &$\frac{1}{4}k_{b}(b\textsuperscript{2}-b\textsubscript{0}\textsuperscript{2})\textsuperscript{2}$& $[\text{kJ/mol/nm\textsuperscript{4}}]$ & $2.306\times 10^{7}$\\
         $b_{\text{OO}}$&$\frac{1}{4}k_{b}(b\textsuperscript{2}-b\textsubscript{0}\textsuperscript{2})\textsuperscript{2}$&$[\text{nm}]$ & 0.1443 \\
         $k_{\text{b(OO)}}$ &$\frac{1}{4}k_{b}(b\textsuperscript{2}-b\textsubscript{0}\textsuperscript{2})\textsuperscript{2}$ &$[\text{kJ/mol/nm\textsuperscript{4}}]$& $6.47\times 10^{6}$ \\
         \vspace{3mm}\\
         $\theta_{\text{HOO}}$ &$\frac{1}{2}k_{\theta}(\cos{\theta}-\cos{\theta}_{0})^{2}$ &$[\text{deg}]$ & 100.36 \\
         $k_{{\theta}{\text{(HOO)}}}$ &$\frac{1}{2}k_{\theta}(\cos{\theta}-\cos{\theta}_{0})^{2}$ &$[\text{kJ/mol]}$& 524.48 \\
         \vspace{3mm}\\
         $C_{0}$ & ${\sum_{n=0}^{5}}~C_{n}(\cos{\psi})^{n}$ &\textendash & 1.995 \\
         $C_{1}$ & ${\sum_{n=0}^{5}}~C_{n}(\cos{\psi})^{n}$ & \textendash& -10.727 \\
         $C_{2}$ & ${\sum_{n=0}^{5}}~C_{n}(\cos{\psi})^{n}$ & \textendash& 15.226 \\
         $C_{3}$ & ${\sum_{n=0}^{5}}~C_{n}(\cos{\psi})^{n}$ & \textendash& -2.244 \\
         $C_{4}$ & ${\sum_{n=0}^{5}}~C_{n}(\cos{\psi})^{n}$ & \textendash& 0.258 \\
         $C_{5}$ & ${\sum_{n=0}^{5}}~C_{n}(\cos{\psi})^{n}$ & \textendash& 0.000 \\
         \vspace{3mm}\\
         $\epsilon_{\text{O}}$ & $4\epsilon_{ij}[(\frac{\sigma_{ij}}{r_{ij}})^{12}-(\frac{\sigma_{ij}}{r_{ij}})^{6}]$ & $[\text{kJ/mol}]$ & 0.743 \\
         $\sigma_{\text{O}}$ & $4\epsilon_{ij}[(\frac{\sigma_{ij}}{r_{ij}})^{12}-(\frac{\sigma_{ij}}{r_{ij}})^{6}]$ &$[\text{nm}]$ & 0.303 \\
         $\epsilon_{\text{H}}$ & $4\epsilon_{ij}[(\frac{\sigma_{ij}}{r_{ij}})^{12}-(\frac{\sigma_{ij}}{r_{ij}})^{6}]$ & $[\text{kJ/mol}]$ & 0.000 \\
         $\sigma_{\text{H}}$ & $4\epsilon_{ij}[(\frac{\sigma_{ij}}{r_{ij}})^{12}-(\frac{\sigma_{ij}}{r_{ij}})^{6}]$ &$[\text{nm}]$ & 0.000 \\
        \vspace{3mm}\\
         $q_{\text{O}}$ & $\frac{1}{4\pi\epsilon_{o}}\frac{q_{i}q_{j}}{r_{ij}}$ & $[e]$ & -0.42 \\
         $q_{\text{H}}$ & $\frac{1}{4\pi\epsilon_{o}}\frac{q_{i}q_{j}}{r_{ij}}$ & $[e]$ & 0.42 \\
         
         \hline
    \end{tabular*}
\end{table*}

\begin{table*} \label{Orabi FF}
\small
  \caption{\ Parameters of potential energy functions corresponding to the bond-stretching, bond-bending, torsional, van der Waals, and electrostatic energies developed by Orabi \& English~\cite{orabi_simple_2018} for H\textsubscript{2}O\textsubscript{2}.}
  \label{tbl:}
  \begin{tabular*}{\textwidth}{@{\extracolsep{\fill}}llll}
    \hline
       Parameter & Functional form &Units & Value\\
    \hline
         $b_{\text{OH}}$ & $\frac{1}{2}k_{b}(b-b_{0})^2 $ & $[\text{nm}]$& 0.0963 \\
         $k_{\text{b(OH)}} $  & $\frac{1}{2}k_{b}(b-b_{0})^2 $ &$[\text{kJ/mol/nm\textsuperscript{2}}]$& 435973\\
         $b_{\text{OO}} $& $\frac{1}{2}k_{b}(b-b_{0})^2 $ & $[\text{nm}]$ & 0.1442 \\
         $k_{\text{b(OO)}} $ & $\frac{1}{2}k_{b}(b-b_{0})^2 $ &$ [\text{kJ/mol/nm\textsuperscript{2}}]$& 238906 \\
         \vspace{3mm}\\
         $\theta_{\text{HOO}}$& $\frac{1}{2}k_{\theta}(\theta-\theta_{0})^2 $ &$[\text{deg}]$ & 99.92 \\
         $k_{\theta{\text{(HOO)}}} $ &$\frac{1}{2}k_{\theta}(\theta-\theta_{0})^2 $ &$ [\text{kJ/mol/rad\textsuperscript{2}}]$& 505.426 \\
         \vspace{3mm}\\
         $\delta$& $k_{\phi}(1+\cos{2\phi-\delta})$ &$[\text{deg}]$  & 0.0 \\
         $k_{\phi{\text{(HOOH)}}} $ & $k_{\phi}(1+\cos{2\phi-\delta})$ &$ [\text{kJ/mol}]$& 8.452 \\
         \vspace{3mm}\\
         $\epsilon_{\text{O}}$ & $4\epsilon_{ij}[(\frac{\sigma_{ij}}{r_{ij}})^{12}-(\frac{\sigma_{ij}}{r_{ij}})^{6}]$ &$ [\text{kJ/mol}]$ & 0.853 \\
         $\sigma_{\text{O}}$& $4\epsilon_{ij}[(\frac{\sigma_{ij}}{r_{ij}})^{12}-(\frac{\sigma_{ij}}{r_{ij}})^{6}]$ &$ [\text{nm}]$ & 0.298 \\
         $\epsilon_{\text{H}}$ & $4\epsilon_{ij}[(\frac{\sigma_{ij}}{r_{ij}})^{12}-(\frac{\sigma_{ij}}{r_{ij}})^{6}]$ &$ [\text{kJ/mol}]$ & 0.192 \\
         $\sigma_{\text{H}}$& $4\epsilon_{ij}[(\frac{\sigma_{ij}}{r_{ij}})^{12}-(\frac{\sigma_{ij}}{r_{ij}})^{6}]$ &$ [\text{nm}]$ & 0.039 \\
         \vspace{3mm}\\
         $q_{\text{O}}$& $\frac{1}{4\pi\epsilon_{o}}\frac{q_{i}q_{j}}{r_{ij}}$ &$ [e]$ & -0.41 \\
         $q_{\text{H}}$ & $\frac{1}{4\pi\epsilon_{o}}\frac{q_{i}q_{j}}{r_{ij}}$ &$ [e]$ & 0.41 \\

         \hline
    \end{tabular*}
\end{table*}


\begin{figure}[!]
    \centering
    \includegraphics[scale=1.8]{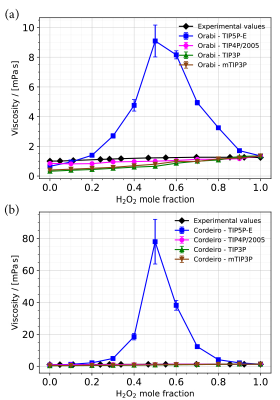}
    \caption{Viscosities of aqueous H\textsubscript{2}O\textsubscript{2} solutions for various mole fractions of H$_2$O$_2$ at 293~K and 1~bar for 
    the (a) Orabi~\cite{orabi_simple_2018} and (b) Cordeiro~\cite{cordeiro_reactive_2014} force fields in combination with the TIP3P~\cite{jorgensen1983comparison}, mTIP3P~\cite{neria1996simulation}, TIP4P/2005~\cite{abascal2005general} and TIP5P-E~\cite{rick2004reoptimization,mahoney2000five} force fields for water. 
    Error bars are estimated based on the standard deviation. The experimental values~\cite{phibbs1951hydrogen} at 293~K and 1~bar are added for comparison.}
    \label{fig:Vvmf_NTIP5P_W_HP}
\end{figure}

\begin{figure}[!]
    \centering
    \includegraphics[scale=0.7]{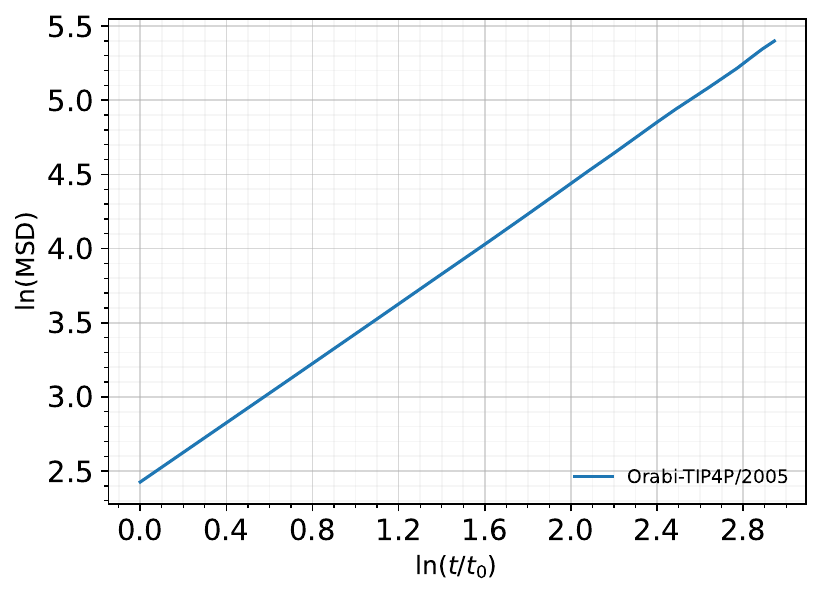}
    \caption{ln(MSD) versus ln(\textit{t$\slash$t\textsubscript{0}}) (\textit{t} stands for time and \textit{t\textsubscript{0}}~=~1~ns) for H\textsubscript{2}O\textsubscript{2} in H$_2$O$_2$ aqueous solution for a H$_2$O$_2$ mole fraction of 0.5 using the Orabi~-~TIP4P/2005 force field combination. The simulation was run for 20~ns in the NVT ensemble from which 1~ns to 20~ns trajectory was used for calculating the self-diffusion coefficients.}
    \label{fig:MSD_05_Orabi_TIP4k5}
\end{figure}

\begin{figure}[!]
    \centering
     \includegraphics[width=0.6\textwidth]{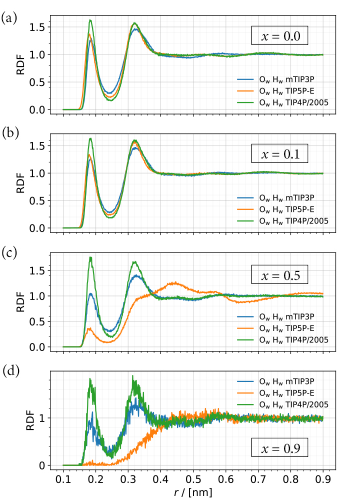}
       \caption{Radial distribution functions~(RDFs) as a function of radial distance, $r$~[nm], for O$_{\text w}$~(O of water)~-~H$_{\text w}$~(H of water) for H\textsubscript{2}O\textsubscript{2} aqueous solutions with $x=$~0.1~(b), $x=$~0.5~(c), and $x=$~0.9~(d), where $x$ is the mole fraction of H\textsubscript{2}O\textsubscript{2}, at 298~K and 1 bar using the Cordeiro~\cite{cordeiro_reactive_2014} force field in combination with the mTIP3P~\cite{neria1996simulation}, TIP5P-E~\cite{rick2004reoptimization,mahoney2000five}, and TIP4P/2005~\cite{abascal2005general} water force fields. The RDF for pure water is plotted in (a).} 
    \label{fig:RDFWater}
\end{figure}

\begin{figure}[!]
    \centering
     \includegraphics[width=0.6\textwidth]{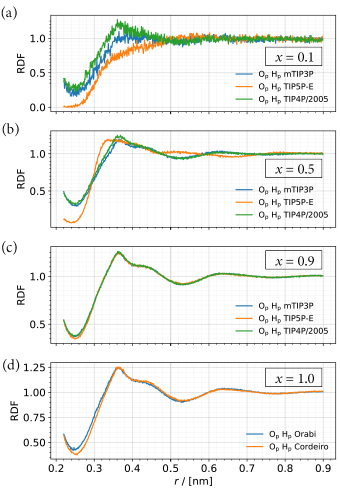}
    \caption{Radial distribution functions~(RDFs) as a function of radial distance, $r$~[nm], for O$_{\text p}$~(O of H\textsubscript{2}O\textsubscript{2})~-~H$_{\text p}$~(H of H\textsubscript{2}O\textsubscript{2}) for H\textsubscript{2}O\textsubscript{2} aqueous solutions with $x=$~0.1~(b), $x=$~0.5~(c), and $x=$~0.9~(d), where $x$ is the mole fraction of H\textsubscript{2}O\textsubscript{2}, at 298~K and 1 bar using the Cordeiro~\cite{cordeiro_reactive_2014} force field in combination with the mTIP3P~\cite{neria1996simulation},  TIP5P-E~\cite{rick2004reoptimization,mahoney2000five}, and TIP4P/2005~\cite{abascal2005general} water force fields. The first peak (at ca. 0.19~nm) was removed to distinguish the differences between the combinations clearly. The RDF for pure H$_2$O$_2$ is plotted in (d) using the Orabi force field and the Cordeiro force field.}
    \label{fig:RDFHP}
\end{figure}

\begin{figure}[!]
    \centering
    \includegraphics[width=1\textwidth]{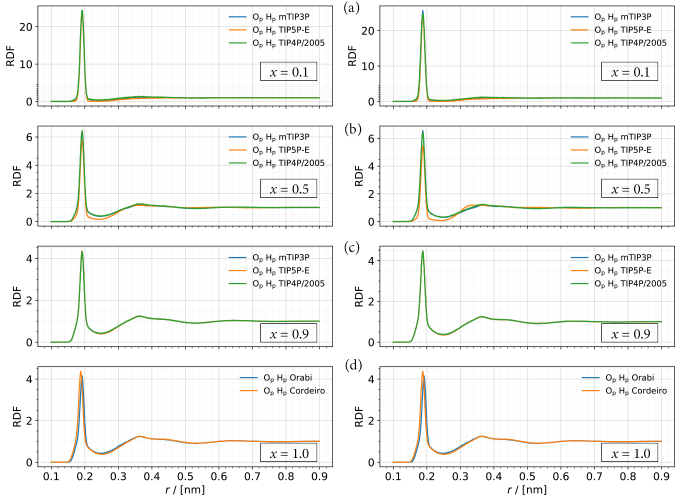}
    \caption{Radial distribution functions~(RDFs) as a function of radial distance, $r$~[nm], for O$_{\text p}$~(O of H\textsubscript{2}O\textsubscript{2})~-~H$_{\text p}$~(H of H\textsubscript{2}O\textsubscript{2}) for H\textsubscript{2}O\textsubscript{2} aqueous solutions with $x=$~0.1~(a), $x=$~0.5~(b), and $x=$~0.9~(c), where $x$ is the mole fraction of H\textsubscript{2}O\textsubscript{2}, at 298~K and 1 bar using the Orabi~\cite{orabi_simple_2018} (left) and Cordeiro~\cite{cordeiro_reactive_2014} (right) force fields in combination with the mTIP3P~\cite{neria1996simulation},  TIP5P-E~\cite{rick2004reoptimization,mahoney2000five}, and TIP4P/2005~\cite{abascal2005general} water force fields. The RDF for pure H$_2$O$_2$ is plotted in (d) using the Cordeiro force field and the Orabi force field.}
    \label{fig:RDFHP}
\end{figure}

\begin{figure*}[!]
    \centering
   \includegraphics[width=1.0\textwidth]{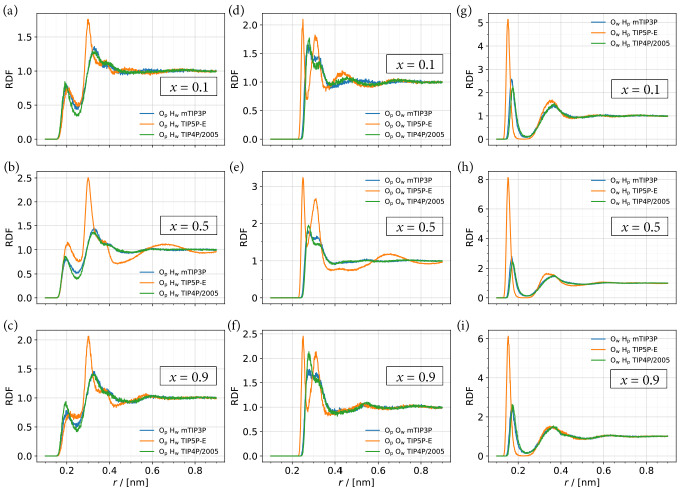}
    \caption{Radial distribution functions~(RDFs) as a function of radial distance, $r$~[nm], for O$_{\text p}$~(O of H$_2$O$_2$) and H$_{\text w}$~(H of water) (a~-~c), O$_{\text w}$~(O of water) and O$_{\text p}$~(O of H$_2$O$_2$) (d~-~f), and O$_{\text w}$~(O of water) and H$_{\text p}$~(H of H$_2$O$_2$) (g~-~i) for systems using the mTIP3P~\cite{neria1996simulation}, TIP5P-E~\cite{rick2004reoptimization,mahoney2000five}, and TIP4P/2005~\cite{abascal2005general} water force fields in combination with the Cordeiro~\cite{cordeiro_reactive_2014} force field for $x=$~0.1, $x=$~0.5, and $x=$~0.9 at 298~K and 1 bar, where $x$ is the mole fraction of H\textsubscript{2}O\textsubscript{2}.}
    \label{fig:RDFCollection}
\end{figure*}

\newpage
\begin{table}[!]
\begin{center}
  \caption{The number of water molecules in the micro solvation shells of H\textsubscript{2}O\textsubscript{2} in aqueous solutions for H$_2$O$_2$ mole fractions of 0.1, 0.5 and 0.9  for the Orabi and Cordeiro force fields in combination with the TIP4P/2005~\cite{abascal2005general}, TIP5P-E~\cite{rick2004reoptimization,mahoney2000five}, TIP3P~\cite{jorgensen1983comparison} and mTIP3P~\cite{neria1996simulation} water force fields. The results are calculated by integrating the RDF for O$_{\text p}$~-~O$_{\text w}$ up to the first minimum ($r_{\textrm{min,\textit{x}}}$), where $x$ is the mole fraction of H\textsubscript{2}O\textsubscript{2} in the system.}
 \label{tbl:Solubility}

  \begin{tabular}{@{\extracolsep{\fill}}llclclc}
    \hline
	Model & 0.1 & $r_{\textrm{min,0.1}}$/[nm] & 0.5 & $r_{\textrm{min,0.5}}$/[nm] & 0.9 &$r_{\textrm{min,0.9}}$/[nm]   \\
		\hline
		Orabi~-~TIP4P/2005 & 2 & 0.31 & 1.26 & 0.31 & 0.25 & 0.31\\
		Orabi~-~TIP5P-E & 1.04 & 0.28 & 0.88 & 0.28 & 0.12 & 0.28\\
  		Orabi~-~TIP3P & 2.09 & 0.31 & 1.17 & 0.31 & 0.24 & 0.31\\
		Orabi~-~mTIP3P & 2.49 & 0.31 & 1.39 & 0.32 & 0.29 & 0.32
    \vspace{3mm}\\
          Cordeiro~-~TIP4P/2005 & 1.77 & 0.3 & 1.15 & 0.31 & 0.23 & 0.31\\
		Cordeiro~-~TIP5P-E & 1.05 & 0.27 & 0.86 & 0.27 & 0.12 & 0.27\\
		Cordeiro~-~TIP3P & 2.17 & 0.31 & 1.06 & 0.3 & 0.24 & 0.31\\
		Cordeiro~-~mTIP3P & 2.26 & 0.31 & 0.98 & 0.3 & 0.22 & 0.31\\
  \hline
  \end{tabular}
  \end{center}
\end{table}
\newpage
\begin{table}
\begin{center}
  \caption{The number of water molecules in the first solvation shells of H\textsubscript{2}O\textsubscript{2} in aqueous solutions for H$_2$O$_2$ mole fractions of 0.1, 0.5 and 0.9  for the Orabi and Cordeiro force fields in combination with the TIP4P/2005~\cite{abascal2005general}, TIP5P-E~\cite{rick2004reoptimization,mahoney2000five}, TIP3P~\cite{jorgensen1983comparison} and mTIP3P~\cite{neria1996simulation} water force fields. The results are calculated by integrating the RDF for O$_{\text p}$~-~O$_{\text w}$ up to the second minimum ($r_{\textrm{min,\textit{x}}}$), where $x$ is the mole fraction of H\textsubscript{2}O\textsubscript{2} in the system.}
 \label{tbl:Solubility}

  \begin{tabular}{@{\extracolsep{\fill}}llclclc}
    \hline
	Model & 0.1 & $r_{\textrm{min,0.1}}$/[nm]& 0.5 & $r_{\textrm{min,0.5}}$/[nm]& 0.9 &$r_{\textrm{min,0.9}}$/[nm]   \\
		\hline
		Orabi~-~TIP4P/2005 & 6.9 & 0.4 & 3.28 & 0.4 & 0.62 & 0.39\\
		Orabi~-~TIP5P-E & 5.89 & 0.38 & 4.24 & 0.39 & 0.58 & 0.37 \\
  		Orabi~-~TIP3P & 6.37 & 0.39 & 3.67 & 0.4 & 0.66 & 0.4 \\
		Orabi~-~mTIP3P & 6.82 & 0.39 & 3.53 & 0.39 & 0.72 & 0.41
    \vspace{3mm}\\
          Cordeiro~-~TIP4P/2005 & 5.6 & 0.38 & 3.21 & 0.39& 0.71 & 0.41\\
		Cordeiro~-~TIP5P-E & 5.07 & 0.36 & 4.32 & 0.38 & 0.55 & 0.37\\
		Cordeiro~-~TIP3P & 5.56 & 0.38 & 3.55 & 0.4 & 0.73 & 0.42 \\
		Cordeiro~-~mTIP3P & 5.55 & 0.38 & 3.59 & 0.4 & 0.68 & 0.4\\
  \hline
  \end{tabular}
  \end{center}
\end{table}
\end{document}